\documentclass[runningheads,a4paper]{llncs}

\usepackage{amssymb}
\usepackage{graphicx}
\usepackage{epstopdf}
\usepackage{subfig}
\usepackage{multirow}
\usepackage{amsmath}
\usepackage{array}
\usepackage{calc}
\usepackage{url}
\usepackage[export]{adjustbox}
\usepackage[belowskip=-10pt,aboveskip=2pt]{caption}
\let\llncssubparagraph\subparagraph
\let\subparagraph\paragraph
\usepackage[compact]{titlesec}
\let\subparagraph\llncssubparagraph
\newcolumntype{C}[1]{>{\centering\let\newline\\\arraybackslash\hspace{0pt}}m{#1}}

\urldef{\mailsa}\path|{alfred.hofmann, ursula.barth, ingrid.haas, frank.holzwarth,|
\urldef{\mailsb}\path|anna.kramer, leonie.kunz, christine.reiss, nicole.sator,|
\urldef{\mailsc}\path|erika.siebert-cole, peter.strasser, lncs}@springer.com|
\newcommand{\keywords}[1]{\par\addvspace\baselineskip
\noindent\keywordname\enspace\ignorespaces#1}

\begin{document}
\parskip 0pt
\mainmatter  

\title{Formalization of Fault Trees in Higher-order Logic: A Deep Embedding Approach\thanks{ The final publication is available at http://link.springer.com}}

\titlerunning{Formalization of Fault Trees in Higher-order Logic}

%
%
\author{Waqar Ahmed %
\and Osman Hasan }
%

\institute{School of Electrical Engineering and Computer Science\\
National University of Sciences and Technology,
Islamabad, Pakistan \\
\email{ \{waqar.ahmad,osman.hasan\}@seecs.nust.edu.pk  }
}

%
%

\maketitle

\begin{abstract}
	
  Fault Tree (FT) is a standard failure modeling technique that has been extensively used to predict reliability, availability and safety of many complex engineering systems. In order to facilitate the formal analysis of FT based analyses, a higher-order-logic formalization of FTs has been recently proposed. However, this formalization is quite limited in terms of handling large systems and transformation of FT models into their corresponding Reliability Block Diagram (RBD) structures, i.e., a frequently used transformation in reliability and availability analyses. In order to overcome these limitations, we present a deep embedding based formalization of FTs.  In particular, the paper presents a formalization of AND, OR and NOT FT gates, which are in turn used to formalize other commonly used FT gates, i.e.,  NAND, NOR, XOR, Inhibit, Comparator and majority Voting, and the formal verification of their failure probability expressions. For illustration purposes, we present a formal failure analysis of a communication gateway software for the next generation air traffic management system.

\end{abstract}

\section{Introduction}

Fault Tree (FT) is used as a standard failure modeling technique in various safety-critical domains, including nuclear power industry, civil aerospace and military systems. It mainly provides a graphical model for analyzing the conditions and
factors causing an undesired top event, i.e., a critical event, which can cause the complete system failure upon its occurrence. The
preceding nodes of the FT are represented by gates, like OR, AND and XOR, which are used to link two or more cause events of a fault in a prescribed manner. Using these FT gates, a FT model of a given system is constructed either on paper or by utilizing graphical editors provided by FT-based computer simulation tools, such as Relia-Soft \cite{Reliasoft_16} and ASENT \cite{ASENT_16}. In the paper-and-pencil proof methods, this obtained FT model is then used for the identification of the Minimal Cut Set (MCS) of failure events that are associated with the components of the given system. This is followed by associating the failure random variables, i.e., exponential or Weibull, to these MCS failure events. The Probabilistic Inclusion-Exclusion (PIE) principle \cite{Trivedi_02} is then used to evaluate the exact probability of failure of the overall system. On the other hand, the FT-based computer tools can be utilized to build a FT model by associating appropriate random variables with each component of the system. The reliability and the failure probability analysis of the complete system is then carried out by using computer arithmetic and numerical techniques on the generated samples from these random variables. However, both these methods cannot ascertain absolute correctness due to their inherent inaccuracy limitations. For instance, paper-and-pencil methods are prone to human errors, especially for large and complex systems, where a FT may consist of 50-130 levels of logic gates  \cite{epstein2005can}. Manually manipulating such a large data makes it quite probable that some of MCS failure events may be overlooked, which would in turn lead to an erroneous  design \cite{epstein2005can}. On the other hand, software tools can efficiently handle the analysis of large FTs but the computational requirements drastically increase as the size of the FT increases.

To overcome the above-mentioned limitations, a higher-order-logic formalization of some basic FT gates and their corresponding failure probability expressions \cite{WAhmed_CICM15} has been recently proposed. However, a major drawback of this formalization is the increase in complexity when analyzing FT of large and complex system. This formalization was primarily based on a shallow embedding approach, where the notion of each FT gate was explicitly defined on an event list and then its corresponding failure probability relationship was verified on the given failure event list. This approach makes the FT gate formalization non-compositional in nature, i.e., the basic FT gates, such as AND, OR and NOT, cannot be used to formalize other FT gates that are usually composed from these basic FT gates. Also, this work \cite{WAhmed_CICM15} utilizes the PIE principle to formally compute the exact failure probability of the given system, which limits its usability for complex system due to the involvement of large number of PIE terms. In the literature, several methods have been used to deal with this inherent complexity issue of the PIE principle. A tractable solution is to transform the given system FT to its equivalent Reliability Block Diagram (RBD) \cite{Bilinton_1992}, which is also a well-known reliability modeling technique. This transformation considerably reduces the analysis complexity  due to the fact that RBD offers closed form expressions compared to a FT, which requires unfolding of all the PIE terms.

In order to overcome the above-mentioned scalability issues of the existing formalization of FT gates \cite{WAhmed_CICM15} and thus broaden the scope of formal FT analysis, we propose a deep embedding approach to formalize the commonly used FT gates, such as AND, OR and NOT. This proposed formalization approach is compositional in nature and can be easily extended to formalize other FT gates, such as NAND, NOR, XOR, Inhibit, Comparator and majority Voting. It also enables us to transform the given system FT model to its equivalent RBD model, without any loss of valuable information. The RBD model can then be formally analyzed using our recently proposed formal reasoning support for RBDs \cite{ahmed2016formalization}.

To illustrate the practical effectiveness of our proposed approach, we present a formal failure analysis of a Next Generation (NextGen) Air Traffic Management (ATM) gateway system, which is primarily used  to enhance the safety and
reliability of air transportation, to improve efficiency in the air transportation and to reduce aviation impact on the environment. The FT of the NextGen ATM gateway, which consists of more than 40 basic failure events including
software, hardware, database update and transmission system is divided into four levels. The formally verified failure probability expressions of individual levels are then used to reason about the failure probability of the overall NextGen system. In addition, we also provide some automated reasoning support for the FT based failure analysis. This automation allows us to automatically simplify the failure expression of the NextGen system from the given values of the failure rates.

\section{Related Work}
\label{sec:related_work}
The COMPASS tool-set \cite{bozzano2009compass} supports the dynamic FT analysis specifically for aerospace systems using the NuSMV and MRMC model checkers. The Interval Temporal Logic (ITS), i.e., a temporal logic that supports first-order logic, has been used, along with the Karlsruhe Interactive Verifier (KIV), for formal FT analysis of a rail-road crossing \cite{ortmeier2007formal}. A deductive method for FT construction, in contrast to the intuitive approach followed in \cite{ortmeier2007formal}, by using the Observational Transition Systems (OTS), is presented in \cite{xiang2004fault}. The formal analysis of this FT is then carried out using CafeOBJ \cite{futatsugi2000cafe}, which is a formal specification language with interactive verification support. However, the scope of these tools is somewhat limited in terms of handling larger systems, due to the inherent state-space explosion problem of model checking. Moreover, either some of these approaches \cite{ortmeier2007formal,xiang2004fault} do not cater for probabilities or if they do cater for them then the computation of probabilities in these methods \cite{bozzano2009compass} involves numerical techniques, which compromises the accuracy of the results.

Leveraging upon the high expressiveness of higher-order logic and the inherent soundness
of theorem proving, Mhamdi's formalized probability theory
\cite{mhamdi_11} has been recently used for the formalization of RBDs \cite{ahmed2016formalization}, including series \cite{WAhmad_CICM14}, parallel
\cite{WAhmed_Wimob15}, parallel-series \cite{WAhmed_Wimob15} and series-parallel \cite{WAhmed_IWIL15}. These formalizations have been used for the
reliability analysis of many applications including simple oil and gas pipelines with serial components \cite{WAhmad_CICM14}, wireless sensor network protocols \cite{WAhmed_Wimob15}
and logistic supply chains \cite{WAhmed_Wimob15}. Similarly, Mhamdi's probability theory have also been used for the formalization of commonly used FT gates, such as AND, OR, NAND, NOR, XOR and NOT, and the PIE principle \cite{WAhmed_CICM15}. In addition, the above-mentioned RBD and FT formalizations have been recently utilized for availability analysis \cite{ahmed2016formal}.  In this paper, we have formalized the FT gates using a deep embedding approach to facilitate the analysis of larger FTs. Besides the existing formalization of FT gates \cite{WAhmed_CICM15}, this paper also provides the formalization of inhibit, 2-bit comparator and Majority voting FT gates. Moreover, we have  combined our existing formalizations of RBDs \cite{WAhmad_CICM14,WAhmed_Wimob15,WAhmed_IWIL15} to make the formal FT based analysis more scalable.

\section{Probability Theory and Fault Trees in HOL}
\label{sec:prob_FT_hol}
Mathematically, a measure space is defined as a triple ($\Omega,\Sigma, \mu$), where
$\Omega$ is a set, called the sample space, $\Sigma$ represents a $\sigma$-algebra of subsets of
$\Omega$, where the subsets are usually referred to as measurable sets, and $\mu$ is a measure with domain
$\Sigma$. A probability space is a measure space ($\Omega,\Sigma, Pr$), such that the measure,
referred to as the probability and denoted by $Pr$, of the sample space is 1. In the HOL4 formalization of probability theory \cite{mhamdi_11}, given a probability space $p$, the functions \texttt{space}, \texttt{subsets} and \texttt{prob} return the corresponding $\Omega$, $\Sigma$ and $Pr$, respectively. This formalization also includes the formal verification of some of the most widely used probability axioms, which play a pivotal role in formal reasoning about reliability properties.

A random variable is a measurable function between a probability space and a measurable space. The measurable functions belong to a special class of functions, which preserves the property that the inverse image of each measurable set  is also measurable. A measurable space refers to a pair ($S,\mathcal{A}$), where $S$ denotes a set and $\mathcal{A}$ represents a nonempty collection of sub-sets of $S$. Now, if $S$ is a set with finite elements, then the corresponding random variable is termed as a discrete random variable otherwise it is called a continuous one.

The cumulative distribution function (CDF) is defined as the probability of the event where a random variable $X$ has a value less than or equal to some value $t$, i.e., $Pr(X \le t)$. This definition characterizes the distribution of both discrete and continuous random variables and has been formalized \cite{WAhmad_CICM14} as follows:

\begin{flushleft}
	\label{CDF_def}
	\vspace{1pt} \texttt{$\vdash$ $\forall$  p X t. CDF p X t = distribution p X \{y | y $\leq$ Normal t\}
	}
\end{flushleft}

\noindent The function \texttt{Normal} takes  a $real$ number as its input and converts it to its corresponding value in the $extended$-$real$ data-type, i.e, it is the $real$ data-type with the inclusion of  positive and negative infinity. The function \texttt{distribution} takes three parameters:  a probability space \texttt{$p:(\alpha \rightarrow bool) \# ((\alpha \rightarrow bool) \rightarrow bool) \# ((\alpha \rightarrow bool) \rightarrow real)$}, a random variable $X: (\alpha \rightarrow extreal)$ and a set of $extended$-$real$ numbers and returns  the probability of the given random variable $X$ acquiring  all the values of the given set in probability space $p$.

The unreliability or the probability of failure $F(t)$ is defined as the probability that a system or component will fail by the time $t$. It can be described in terms of CDF, known as the failure distribution function, if the random variable $X$ represent a time-to-failure of the component. This time-to-failure random variable $X$ usually exhibits the exponential or Weibull distribution.

The notion of mutual independence of $n$ random variables is a major requirement for reasoning about the failure analysis of most of the FT gates. According to this notion, a list of $n$ events are mutual independent if and only if for each set of $k$ events, such that $\mathit(1 \le k \le n)$, we have:
\begin{equation}\label{eq1:mutual_indep}
Pr(\bigcap_{i=1}^{k}A_i) = \prod_{i=1}^{k} Pr(A_i)
\end{equation}

It is important to note that mutual independence is a much stronger property compared to pairwise independence \cite{Trivedi_02}, which ensures independence between two events only. On the other hand, mutual independence makes sure that any subset of events are independent with each other. Also, we can verify many interesting properties of independence using the mutual independence property. For instance, given a list of mutually independent events, say $L$, we can verify that an element $h \in L$ is independent with the list $L - [h]$ representing the list $L$ without element $h$.

The mutual independence concept is formalized in HOL4 as follows \cite{WAhmad_CICM14}:
\vspace{5mm}
\begin{flushleft}
	\label{mutual_indep_def}
	\vspace{1pt} \small{\texttt{$\vdash$ $\forall$ p (L:$\alpha \rightarrow bool$). mutual\_indep p L  =
			$\forall$  L1 \vspace{1pt} (n:num). PERM L L1 $\wedge$\\
			\ \   1 $\leq$ n $\wedge$ n $\leq$ LENGTH L $\Rightarrow$ \\
			\ \ \   prob p (inter\_list p (TAKE n L1)) = list\_prod (list\_prob p (TAKE n L1))
		}}
	\end{flushleft}
	\noindent  The function \texttt{mutual\_indep}  accepts a list of events $L$ and probability space $p$ and returns $True$ if the events in the given list are mutually independent in the probability space $p$.  The predicate \texttt{PERM} ensures that its two lists as its arguments  form a permutation of one another. The function \texttt{LENGTH} returns the length of the given list. The function \texttt{TAKE} returns the first $n$ elements of its argument list as a list. The  function \texttt{inter\_list} performs the intersection of all the sets in its argument list of sets  and returns the probability space if the given list of sets is empty. The function \texttt{list\_prob} takes a list of events and returns a list of probabilities associated with the events in the given list of events in the given probability space. Finally, the function \texttt{list\_prod} recursively multiplies all the elements in the given list of real numbers. Using these functions, the function \texttt{mutual\_indep} models the mutual independence condition such that for $n$ events taken from any permutation of the given list $L$, Equation (\ref{eq1:mutual_indep}) holds.

\subsection{Formalization of Fault Tree Gates}
\label{sec:form_FTGates}

The proposed formalization is primarily based on defining a new polymorphic datatype \textit{gate} that encodes the notion of AND, OR and NOT FT gates. Then a semantic function is defined on that \textit{gate} datatype yielding an event for the corresponding FT gate. This semantic function allows us to verify the generic failure probability expressions of the FT gates by utilizing the underlying probability theory within the sound core of the HOL4 theorem prover. Such a deep embedding considerably simplifies the FT gate modeling approach, compared to our previous work \cite{WAhmed_CICM15} (shallow embedding), and also enables us to develop a framework that can deal with arbitrary levels of FTs, which can be used to cater for a wide variety of real-world failure analysis problems.

We start the formalization process by type abbreviating the notion of event, which is essentially a set of observations with type \texttt{'a->bool} as follows:

\begin{flushleft}
	\texttt{\small  type\_abbrev ("event" , ``:'a ->bool'')}
\end{flushleft}

\noindent We then define a recursive datatype $gate$ in the HOL4 system as follows:
\begin{flushleft}
	\texttt{\small   Hol\_datatype `gate = AND of gate list | OR of gate list | NOT of gate |\\
	\	\qquad \qquad \qquad \qquad  \qquad  atomic of 'a  event`}
\end{flushleft}

\noindent The type constructors \texttt{AND} and \texttt{OR} recursively function on \textit{gate}-typed lists and the type constructor \texttt{NOT} operates on \textit{gate}-type variable. The type constructor \texttt{atomic} is basically a typecasting operator between \textit{event} and \textit{gate}-typed variables. These type constructors allow us to encode the notion of all the basic FT gates.

We define a semantic function \texttt{$FTree:\alpha \hspace{1mm} event \hspace{1mm} \# \hspace{1mm} \alpha \hspace{1mm} event \hspace{1mm} event \hspace{1mm} \# \hspace{1mm} (\alpha \hspace{1mm} event \rightarrow real) \hspace{1mm} \rightarrow \hspace{1mm} \alpha \hspace{1mm} gate \hspace{1mm} \rightarrow \hspace{1mm} \alpha \hspace{1mm} event$} over the above-defined \textit{gate} datatype that can yield the corresponding event from the given FT gate as follows:

\begin{flushleft}
	\texttt{\bf{Definition 1: }}
	\label{rbd_struct}
	\small
	\vspace{1pt} \texttt{$\vdash$
		($\forall$ p. FTree p (AND []) = p\_space p) $\wedge$\\
		($\forall$ xs x p.
		FTree p (AND (x::xs)) = FTree p x $\cap$ FTree p (AND xs)) $\wedge$\\
		($\forall$ p. FTree p (OR []) = \{\}) $\wedge$\\
		($\forall$ xs x p. FTree p (OR (x::xs)) = FTree p x $\cup$ FTree p (OR xs))  $\wedge$ \\
		($\forall$ p a. FTree p (NOT a) =  p\_space p DIFF FTree p a) $\wedge$\\
		($\forall$ p a. FTree p (atomic a) = a)
	}
\end{flushleft}

\noindent The above function decodes the semantic embedding of a FT by yielding a corresponding failure event, which can then be used to determine the failure probability of a given FT. The function \texttt{FTree} takes a list of type \emph{gate},  identified by a type constructor \texttt{AND}, and returns the whole probability space if the given list is empty and otherwise returns the intersection of the events that are obtained after applying the function \texttt{FTree} on each element of the given list in order to model the AND FT gate behaviour. Similarly, to model the behaviour of the OR FT gate, the function \texttt{FTree} operates on a list of datatype \emph{gate}, encoded by a type constructor $\texttt{OR}$. It then returns the union of the events after applying the function \texttt{FTree} on each element of the given list or an empty set if the given list is empty. The function \texttt{FTree} takes a type constructor \texttt{NOT} and returns the complement of the failure event obtained from the function \texttt{FTree}. The function \texttt{FTree} returns the failure event using the type constructor \texttt{atomic}.

If the occurrence of the failure event at the output is caused by the occurrence of all the input failure events then this kind of behavior can be modeled by using the AND FT gate. The failure probability expression of the AND FT gate can be expressed mathematically as follows:

\begin{equation}\label{eq:and_gate}
F_{AND\_gate}(t) = Pr (\bigcap_{i=2}^{N}A_{i}(t))
= \prod_{i=2}^{N}F_{i}(t)
\end{equation}

\noindent Using Definition 1, we can verify the above equation in HOL4 as follows:

\begin{flushleft}
	\small {\texttt{\bf{Theorem 1: }}} \label{AND_FT_THM}
	\vspace{1pt} \small{\texttt{$\vdash$ $\forall$ p L.
			prob\_space p $\wedge$
			\\ ($\forall$x'. MEM x' L $\Rightarrow$ x' $\in$ events p) $\wedge$ 2 $\le$ LENGTH L $\wedge$\\
			mutual\_indep p L $\Rightarrow$\\
			\ (prob p (FTree p (AND (gate\_list L))) = list\_prod (list\_prob p L))
		}}
	\end{flushleft}
	
	\noindent The first two assumptions, in Theorem 1, ensures that $p$ is a valid probability space and each element of a given event list $L$ must be in event space $p$
	based on the probability theory in HOL4  \cite{mhamdi_11}. The function \small{\texttt{MEM}} finds an element in a given list and returns false, if a match does not occur. The next two assumptions guarantee that the list of events $L$, representing the failure probability of individual components, must have at least two events and the failure events are mutually independent. The conclusion of the theorem represents Equation (\ref{eq:and_gate}). The function \small{\texttt{gate\_list}} generates a list of type $gate$ by mapping the function \texttt{atomic} to each element of the given event list $L$ to make it consistent with the assumptions of Theorem 1. It can be formalized in HOL4 as: \small{\texttt{$\forall$ L. gate\_list L = MAP ($\lambda$a. atomic a) L}}

	\noindent The proof of Theorem 1 is primarily based on a mutual independence property and some fundamental axioms of probability theory.
	
In the OR FT gate, the occurrence of the output failure event depends upon the occurrence of any one of its input failure event. Mathematically, the failure probability of an OR FT gate can be expressed as:

\begin{equation}\label{eq4:OR_gate}
      F_{OR\_gate}(t) = Pr (\bigcup_{i=2}^{N}A_{i}(t))
      = 1 - \prod_{i=2}^{N}(1 - F_{i}(t))
 \end{equation}

By following the approach, used in Theorem 1, we can formally verify the failure probability expression OR FT gate, given in Equation (\ref{eq4:OR_gate}), in HOL4:

\begin{flushleft}
	\small{\texttt{\bf{Theorem 2: }}} \label{parallel_connected_system_THM}
	\vspace{1pt} \small{\texttt{$\vdash$ $\forall$ p L.
			prob\_space p $\wedge$
			2 $\le$ LENGTH L $\wedge$\\ ($\forall$x'. MEM x' L $\Rightarrow$ x' $\in$ events p) $\wedge$
			mutual\_indep p L $\Rightarrow$\\
			\  (prob p (FTree p (OR (gate\_list L))) =\\
			\ \ 1 - list\_prod (one\_minus\_list (list\_prob p L)))
		}}
	\end{flushleft}
	
	\noindent The above theorem is verified under the same assumptions as Theorem 1. The conclusion of the theorem represents Equation (\ref{eq4:OR_gate}) where, the function \small{\texttt{one\_minus\_list}} accepts a list of $real$ numbers $[x1, x2, x3, \cdots, xn]$ and returns the list of $real$ numbers such that each element of this list is 1 minus the corresponding element of the given list, i.e., \small{$[1-x1, 1-x2, 1-x3, \cdots, 1-xn]$}.
	
The NOT FT gate can be used in conjunction with the AND and OR FT gates to formalize other FT gates. The formalization of these gates is given in Table \ref{table:FT_gate_def}. The NAND FT gate, represented by the function \small{\texttt{NAND\_FT\_gate}} in Table \ref{table:FT_gate_def}, models the behavior of the occurrence of an output failure event when at least one of the failure events at its input does not occur. This type of gate is used in FTs when the non-occurrence of the failure event in conjunction with the other failure events causes the top failure event to occur. This behavior can be expressed as the intersection of complementary and normal events, where the complementary events model the non-occurring failure events and the normal events model the occurring failure events. The output failure event occurs in the 2-input XOR FT gate if only one, and not both, of its input failure events occur. The inhibit FT gate produces an output failure event only if the conditional event occurs at the same time when the input failure event occurs. The HOL4 function \small{\texttt{inhibit\_FT\_gate}}, given in Table \ref{table:FT_gate_def}, models the behavior of a 2-input inhibit FT gate by composing the type constructors \textit{AND}, \textit{OR} and \textit{NOT}. In the comparator FT gate, the output failure event occurs if all the failure events at its input occur or if all of the them do not occur.  In the majority voting gate, the output failure event occurs if at least $m$ out of $n$ input failure events occurs. This behaviour can be modeled by utilizing the concept of binomial trials, which are used to find the chances of at least \textit{m} success in \textit{n} trials. The function \small{\texttt{major\_voting\_FT\_gate}} accepts a probability space $p$, a binomial random variable $X$ and two variables, $m$ and $n$, which represent the number of successes and total number of trials, respectively. It then returns the union of the corresponding events that are associated with the binomial random variable $X$, which takes values from the set \small{\{x $|$ k $\leq$ x $\wedge$ x $<$ SUC n\}}. The function \texttt{IMAGE} takes a function \textit{f} and an arbitrary domain set and returns a range set by applying the function \textit{f} to all the elements of the given domain set. The function \texttt{BIGUNION} returns the union of all the element of given set of sets.
	
	 \begin{table}[!htb]
	 	\centering
	 	\caption{HOL4 Formalization of Fault Tree Gates}
	 	\scalebox{0.8}{
	 		\begin{tabular}{|l |l|}
	 			\hline
	 			FT Gates & Formalization  \\
	 			\hline
	 			\hline
	 			\includegraphics[valign=c,scale=0.3,trim={1.5cm 0.7cm 0 0.7cm},clip]{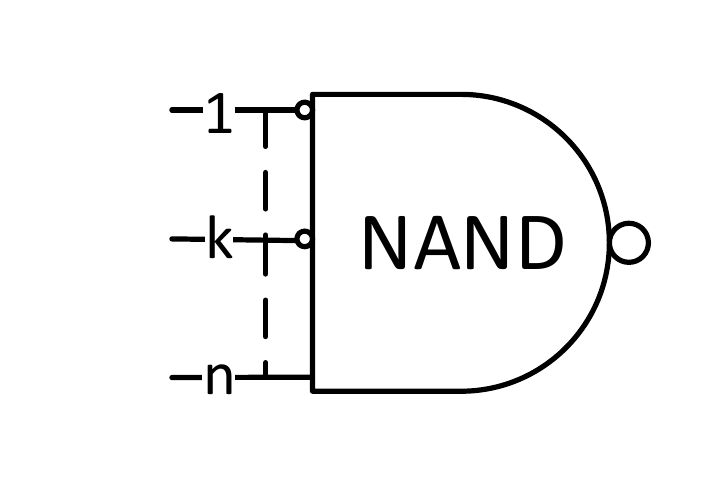} & 	
                $\!\begin{aligned}[c]
                    &\small{\texttt{$\vdash$ $\forall$ p L1 L2. NAND\_FT\_gate p L1 L2 = }}\\
                    & \qquad \qquad \small{ \texttt{FTree p (AND (gate\_list (compl\_list p L1 ++ L2)))}
	 			}\end{aligned}$ \\
	 			\hline
	 			\includegraphics[valign=c,scale=0.3,trim={1.5cm 0.6cm 0 0.6cm},clip]{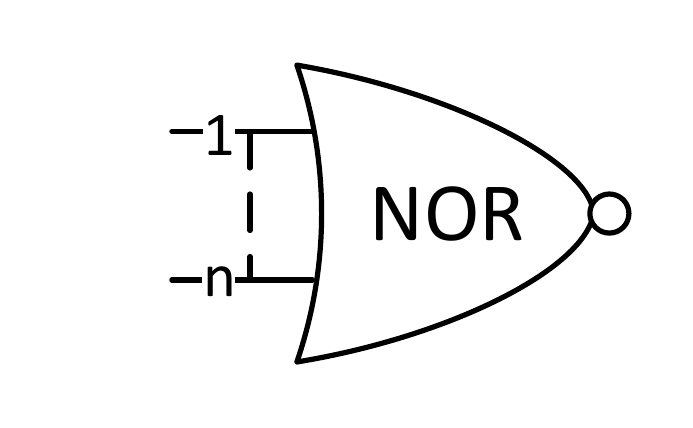}&
                $\!\begin{aligned}[c]
                    \small{\texttt{$\vdash$ $\forall$ p L. NOR\_FT\_gate p L = FTree p (NOT (OR (gate\_list L)))}
	 			}\end{aligned}$ \\
	 			\hline		
	 			\includegraphics[valign=c,scale=0.3,trim={1.5cm 0.6cm 0 0.6cm},clip]{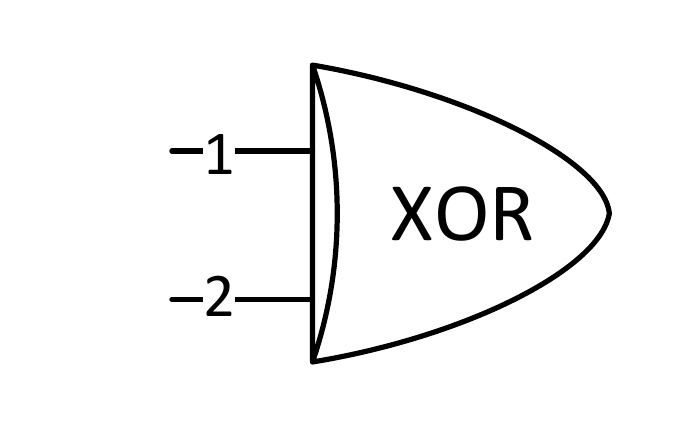} &
                $\!\begin{aligned}[c]
                    & \small{\texttt{$\vdash$ $\forall$ p A B. XOR\_FT\_gate p A B =}}\\
                    & \qquad \qquad \small{\texttt{FTree p (OR [AND [NOT A; B]; AND [A; NOT B]])
	 			}}\end{aligned}$ \\
	 				\hline					
	 			\includegraphics[valign=c,scale=0.28,trim={0 0.6cm 1cm 0.6cm},clip]{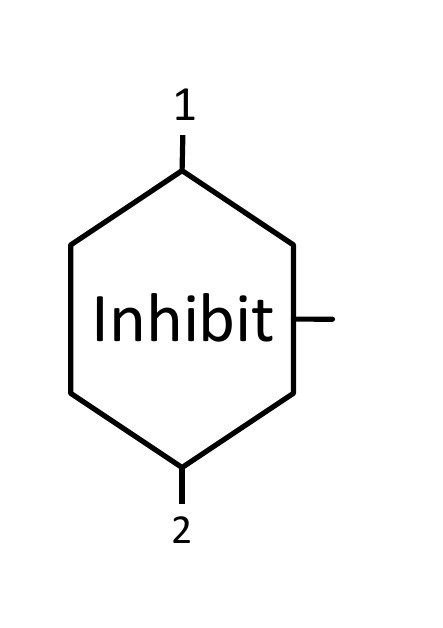} &
                $\!\begin{aligned}[c]
                    & \small{\texttt{$\vdash$ $\forall$ p A B C. inhibit\_FT\_gate p A B C =}}\\
                    & \qquad \qquad \small{\texttt{FTree p (AND [OR [A; B]; NOT C]])
	 			}}\end{aligned}$ \\
	 					\hline							
	 			\includegraphics[valign=c,scale=0.3,trim={1.5cm 0.6cm 0 0.6cm},clip]{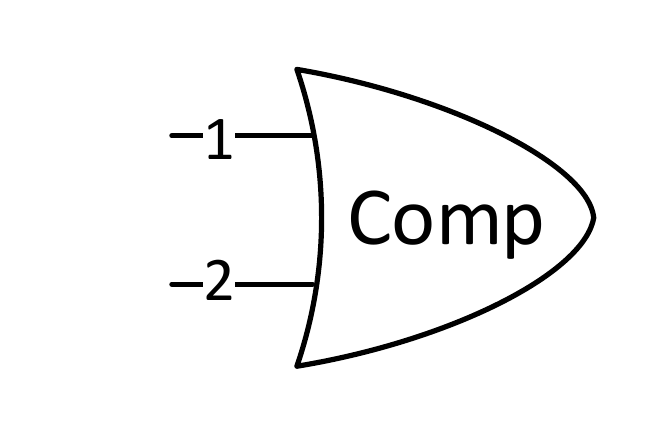} &
                $\!\begin{aligned}[c]
                    & \small{\texttt{$\vdash$ $\forall$ p A B. comp\_FT\_gate p A B =}}\\
                    & \qquad \qquad \small{\texttt{FTree p (OR [AND [A; B]; NOR\_FT\_gate p [A; B]])
	 			}}\end{aligned}$ \\
	 				\hline
	 			\includegraphics[valign=c,scale=0.3,trim={1.5cm 0 0 0},clip]{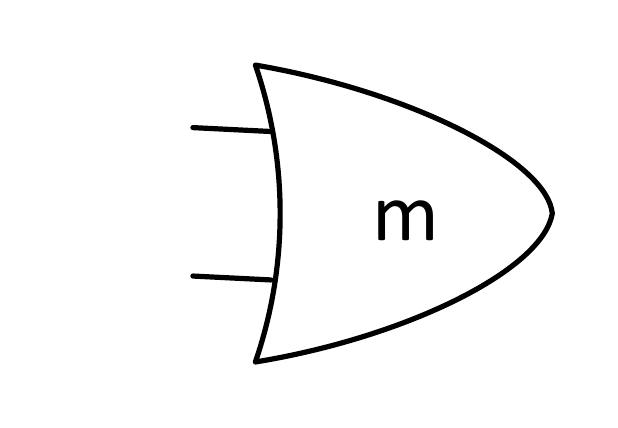} &
                $\!\begin{aligned}[c]
                    & \small{\texttt{$\vdash$ $\forall$ p X m n. major\_voting\_FT\_gate p X m n =}}\\
                    &  \ \small{\texttt{BIGUNION (IMAGE ($\lambda$x. PREIMAGE X \{Normal (\&x)\} $\cap$ p\_space p)	}} \\
                    & \qquad \qquad \small{\texttt{\{x | k $\leq$ x $\wedge$ x  < SUC n\})
	 		    }}\end{aligned}$ \\
	 				\hline
	 \end{tabular}}\label{table:FT_gate_def}
	\end{table}

	 The verification of the corresponding failure probability expressions, of the above-mentioned FT gates, is presented in Table \ref{FT_exp_table}. These expressions are verified under the same assumptions as the ones used for Theorems 1 and 2.	However, some additional provisos are required for the verification of majority voting gate as follows: (i) \texttt{prob\_space} ensures that $p$ is a valid probability space; (ii) \texttt{m $\leq$ n} makes sure that the number of successes of trails $m$ must be less than or equal the total number of trials $n$; (iii) \texttt{($\lambda$x. PREIMAGE X {Normal(\&x)} $\cap$ p\_space p) $\in$ ((count (SUC n)) $\rightarrow$ events p)} ensures that all the corresponding events that are associated with the binomial random variable $X$ are drawn from the events space $p$; and (iv) \texttt{($\forall$x. distribution p X \{Normal (\&x)\} = (\&binomial n x)*(F pow x)*(1 - F) pow (n-x))} guarantees that the random variable $X$ is exhibiting the binomial distribution.		
							
\begin{table}[!htb]
		\centering
		\caption{Probability of Failures of Fault Tree Gates}
		\scalebox{0.78}{
	\begin{tabular}{|l|l|}
				\hline
				Mathmatical Expressions & Theorem's Conclusion \\
				\hline
				\hline
						$\!\begin{aligned}[c]
							F_{NAND}(t)
                            & =  Pr (\bigcap_{i=2}^{k}\overline A_{i}(t) \cap \bigcap_{j=k}^{N}A_{i}(t)) \\
                            &= \prod_{i=2}^{k}(1 - F_{i}(t)) *\prod_{j=k}^{N}(F_{j}(t))
                        \end{aligned}$  &
                        $\!\begin{aligned}[c]
                            & \small{\texttt{$\vdash$ $\forall$ p L1 L2. (prob p (NAND\_FT\_gate p L1 L2)  =}}\\
                            & \ \small{\texttt{list\_prod ((list\_prob p (compl\_list p L1))) *}}\\
                            & \small{\texttt{ list\_prod (list\_prob p L2))}}
                        \end{aligned}$ \\
								\hline
						$\!\begin{aligned}[c]
							F_{NOR}(t)
							& = 1 - F_{OR}(t)  = \prod_{i=2}^{N}(1 - F_{i}(t))
						\end{aligned}$  &
						$\!\begin{aligned}[c]
                            & \small{\texttt{$\vdash$ $\forall$ p L. (prob p (NOR\_FT\_gate p L)  =}} \\
                            & \ \  \small{\texttt{list\_prod (one\_minus\_list (list\_prob p L)))}}
                        \end{aligned}$ \\
									\hline
						$\!\begin{aligned}[c]
						  F_{XOR}(t)
                            &= Pr(\overline{A}(t)B(t) \cup A(t)\overline{B}(t)) \\
                            &= (1- F_{A}(t))F_{B}(t) +\\
                            & \ \ \ \    F_{A}(t)(1- F_{B}(t))
                        \end{aligned}$  &
                        $\!\begin{aligned}[c]
                            & \texttt{$\vdash$ $\forall$p A B. prob\_space p $\wedge$ A $\in$ events p $\wedge$ B $\in$ events p}\\
                            & \texttt{(prob p (XOR\_FT\_gate p (atomic A) (atomic B)  =}\\
                            & \texttt{(1- prob p A)*prob p B + prob p A*(1 - prob p B)}
                        \end{aligned}$ \\
								\hline
	                   $\!\begin{aligned}[c]
	                       F_{inhibit}(t)
                            &= Pr((A(t) \cup B(t)) \cap \overline{C(t)}) \\
                            &= (1- (1 - F_{A}(t))*\\ & \ \ \ \ (1 - F_{B}(t)))* (1- F_{C}(t))
                        \end{aligned}$  &
                        $\!\begin{aligned}[c]
                            & \texttt{$\vdash$ $\forall$p A B C. }\\
                            & \small\texttt{(prob p}\\
                            & \ \ \texttt{(inhibit\_FT\_gate p (atomic A) (atomic B) (atomic C)  =}\\
                            & \texttt{(1 - (1 - prob p A) * (1 - prob p B))*(1 - prob p C)}
                        \end{aligned}$ \\
								\hline
	                   $\!\begin{aligned}[c]
	                       F_{comp}(t)
                            &= Pr((A(t) \cap B(t)) \cup \overline{(A(t) \cup B(t))}) \\
                            &= (1- (1 - F_{A}(t)F_{B}(t))*\\
                            & \ \ \ \ (1 - (1 - F_{A}(t))* (1- F_{B}(t)))
                        \end{aligned}$  &
                        $\!\begin{aligned}[c]
                            & \texttt{$\vdash$ $\forall$p A B C. }\\
                            & \small\texttt{(prob p (comp\_FT\_gate p (atomic A) (atomic B)  =}\\
                            & \texttt{(1 - (1 - prob p A * prob p B)*}\\
                            & \ \texttt{(1 - (1 - prob p A)*(1- prob p B))}
                        \end{aligned}$ \\
	                   \hline
		               $\!\begin{aligned}[c]
		                  F_{m|n}(t)
                            &= Pr (\bigcup_{i=k}^{n}\{\textit{exactly \textit{i} components are}\\
                            & \qquad \qquad \textit{functioning properly}\})\\
		                    & = ‎\sum_{i=m}^{n} (\dbinom{n}{m} F^{i} (1 - F)^{n -1})
                        \end{aligned}$  &
                        $\!\begin{aligned}[c]
                            & \texttt{$\vdash$ $\forall$p n k X F }\\
                            & \small\texttt{(prob p (major\_voting\_FT\_gate p X m n) =}\\
                            & \texttt{ sum (m, SUC n - m)}\\
                            & \ \texttt{($\lambda$x. (\&binomial n x)*(F pow x)* (1- F) pow (n-x)))}
                        \end{aligned}$ \\
		                  \hline
    \end{tabular}}\label{FT_exp_table}
\end{table}
\subsection{Formalization of Probabilistic Inclusion-Exclusion Principle }							
 In FT analysis, firstly all the basic  failure events are identified that can cause the occurrence of the system top failure event. These failure events are then combined to model the overall fault behavior of the given system by using the fault gates. These combinations of basic failure events, called cut sets, are then reduced to  minimal cut sets (MCS) by using some set-theory rules, such as idempotent, associative and commutative. Then, the Probabilistic Inclusion Exclusion (PIE) principle is used to evaluate the overall failure probability of the given system based on the MCS events. According to the PIE principle, if $A_{i}$ represents the $i^{th}$ basic failure event or a combination of failure events then the overall failure probability of the given system can be expressed as follows:
 \small
 \begin{equation}\label{PIE}
 \mathbb{P} (\bigcup_{i=1}^n A_i)  = \sum_{t \neq \{\}, t\subseteq\{1,2,\ldots,n\}}(-1)^{|t|+1} \mathbb{P} (\bigcap_{j\in t} A_j)
 \end{equation}
 \normalsize

 \noindent The above equation has been formally verified in HOL as follows \cite{WAhmed_CICM15}:
 \begin{flushleft}
 	\small{\texttt{\bf{Theorem 3: }}} \label{PIE_THM}
 	\vspace{1pt} \small{\texttt{$\vdash$ $\forall$ p L. prob\_space p $\wedge$ ($\forall$ x. MEM x L $\Rightarrow$ x $\in$ events p) $\Rightarrow$ \\
 			\      (prob p (union\_list L) = \\  \ \
 			sum\_set \{t | t  $ \subseteq $ set L $ \wedge $ t $ \neq $ \{\} \}\\ \qquad \qquad
 			($ \lambda $t. -1 pow (CARD t + 1) * prob p (BIGINTER t)))
 		}}
 	\end{flushleft}
 	
 	\noindent The assumptions of the above theorem are the same as the ones used in Theorem 1. The function \texttt{sum\_set} takes an arbitrary set $s$ with element of type $\alpha$ and a real-valued function $f$ and recursively sums the return values of the function $f$, when applied on each element of the given set $s$. In the above theorem, the set $s$ is represented by the term $\{x|C(x)\}$ that contains all the values of $x$, which satisfy condition $C$. Whereas, the $\lambda$ abstraction function \texttt{($ \lambda $t. -1 pow (CARD t + 1) * prob p (BIGINTER t))} models $(-1)^{|t|+1} \mathbb{P} (\bigcap_{j\in t} A_j)$, such that the functions \texttt{CARD} and \texttt{BIGINTER} return the number of elements and the intersection of all the elements of the given set, respectively.
 	
 \subsection{Formalization of Reliability Block Diagrams}	
 	 Transformation of a system FT to its equivalent reliability block diagram (RBD) has been proposed as a viable solution to reduce the complexity associated with finding the failure probability of large systems \cite{kuykendall2002systems}. The proposed deep embedding based formalization of FT gates allows the establishment of this link and thus we have used the existing formalization of RBDs \cite{ahmed2016formalization} to make the formal analysis of FTs more scalable.  In this paper, we only describe the formalization of the parallel-series RBD configuration because it is required to conduct the formal failure analysis of ASN gateway system, described in the next section.
\begin{figure}
  \centering
  \includegraphics[width=0.35\textwidth]{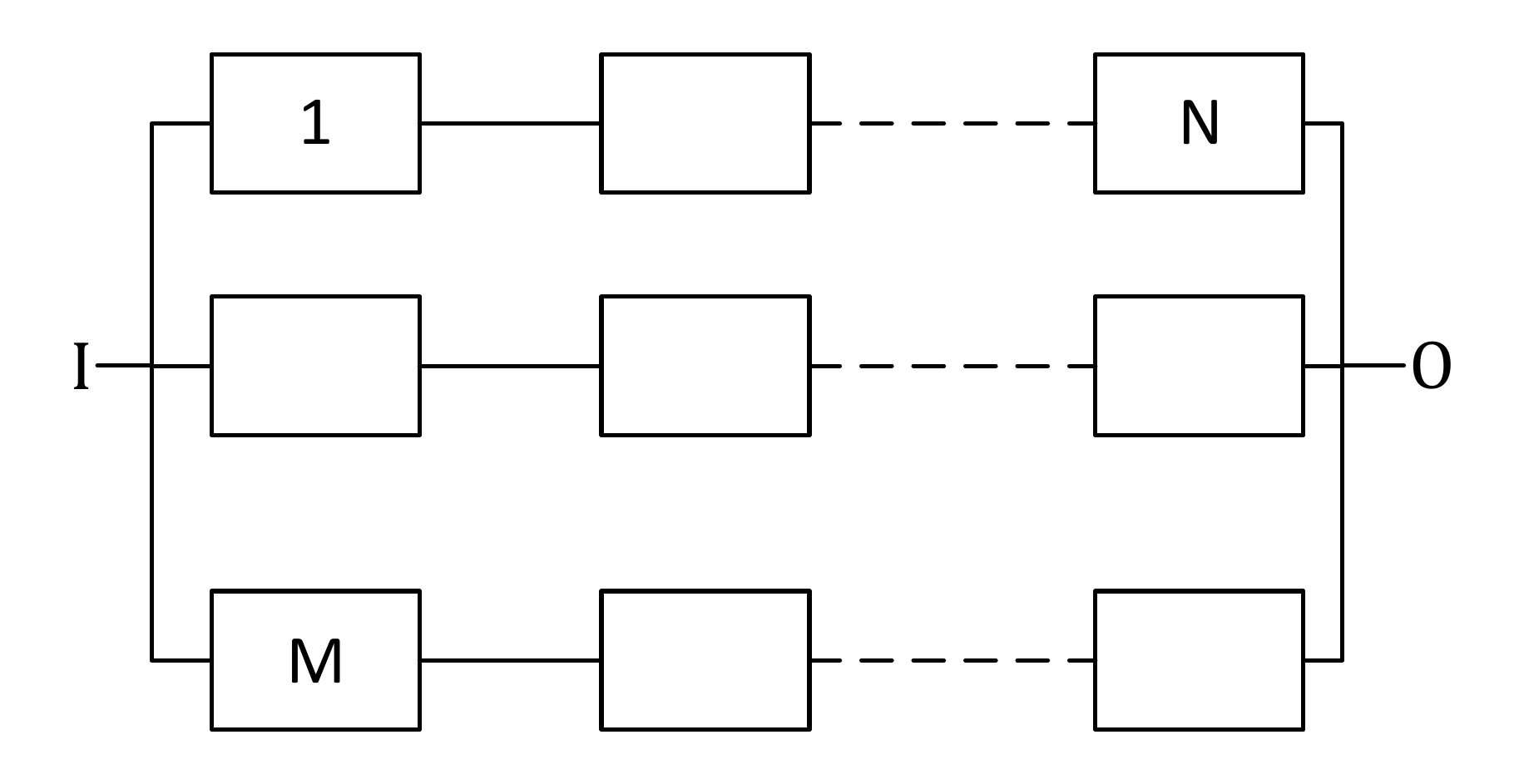}
  \caption{Parallel-Series Reliability Block Diagrams}\label{series_parallel_RBD}
\end{figure}

In a parallel-series RBD configuration, as shown in Fig. \ref{series_parallel_RBD}, the reserved \emph{subsystems} are connected serially and it can be considered as the nested form of series RBD in a parallel RBD configuration. If $A_{ij}(t)$ is the event corresponding to the reliability of the $j^{th}$ component connected in a  $i^{th}$ subsystem at time $t$, then parallel-series RBD configuration can be expressed as:

\begin{equation}\label{eq7:parallel-seriesRBD}
\small R_{parallel-series}(t) = Pr (\bigcup_{i=1}^{M} \bigcap_{j=1}^{N} A_{ij}(t))= 1- \prod_{i=1}^{M}(1 - \prod_{j=1}^{N} (R_{ij}(t)))
\end{equation}

\normalsize
The HOL4 formalization of the above equation is as follows \cite{ahmed2016formalization}:

\begin{flushleft}
	\small{\texttt{\bf{Theorem 4: }} \label{parallel_series_connected_system_THM}
		\vspace{1pt} \texttt{$\vdash$ $\forall$ p L.
			prob\_space p $\wedge$
			($\forall$z. MEM z L $\Rightarrow$  $\backsim$NULL z) $\wedge$\\
			($\forall$x'. MEM x' (FLAT L) $\Rightarrow$  x' $\in$ events p) $\wedge$ \\
			\ mutual\_indep p (FLAT L) $\Rightarrow$\\
			\  (prob p (rbd\_struct p ((parallel of
			($\lambda$a. series (rbd\_list a))) L)) =\\
			\     (1 - list\_prod (one\_minus\_list) of ($\lambda$a. list\_prod (list\_prob p a))) L)
		}}
	\end{flushleft}

	\noindent where the function \texttt{rbd\_struct} is defined on a recursive datatype $rbd$ and can take any combination of type constructors \texttt{series} and \texttt{parallel}. It then yields the corresponding event of the given RBD configuration constituted by these type-constructors. The function \texttt{rbd\_list} serves similar functionality as that of the function \texttt{gate\_list}. The assumptions are quite similar to the ones used for Theorems 1 and 2. The conclusion models Equation (\ref{eq7:parallel-seriesRBD}) and the infixr function \texttt{of} connects two $rbd$ type-constructors by using the HOL4 \texttt{MAP} function.

\section{Formalization of the NextGen ASN Gateway System}
NextGen is supported by the nation-wide Aviation Simulation Network (ASN), which is an environment including simulated and human-in-the-loop (HIL) real-life components, e.g., pilots and air traffic controllers. The Real Time Distributed Simulation (RTDS) application suite \cite{torngren1998fundamentals} is used to facilitate the ASN by providing low and medium fidelity en-route simulation capabilities. An ASN gateway software system acts as an intermediary between RTDS and ASN by providing logic for data translation, two-way communication and transfer messages among them. The overall NextGen ASN gateway FT can be viewed as a four level FT \cite{kornecki2013fault}. The first or top level of the ASN gateway FT models an aviation accident caused by the lack of appropriate control, equipment, internal and external malfunctions. The internal failure event opens up to a second level of the ASN gateway FT, which comprises of failures related to the flight function mishap and transmissions. The flight mishap failure is caused by the failure of the Auto Pilot (AP) or Flight Director (FD) along with the failure not mitigated in time (FF1). The Transmission failure event captures the failure events due to data/message not correctly transmitted (A), failure to display (NotShown), and not performing transmission in a timely manner (RT). The third level of the ASN gateway FT is composed of several sub-FTs, given in Table \ref{table:ASN_FT}, representing the RT and failure event A. The RT failure event occurs if the delay is too long for the transmission to meet its deadline (Time) and a latency problem occurs related to either the application (AL), serialization (SL), propagation delay (PD) or any other relevant sources. Similarly, the failure event A represents a failure to correctly transmit a message and consists of two events. i.e., B1: failure to
transfer a message from ASN to RTDS and B2: failure to transfer a message from RTDS to ASN of the communication link. The FT of the events B1 and B2 are given at the fourth level of the ASN gateway FT \cite{kornecki2013fault}. The overall ASN gateway FT consists of 47 basic failure events that are related to messages transmission failures, propagation delays, software and hardware equipment failures, database update failures and human mistakes.

\subsection{Formal Fault Tree Models for ASN Gateway System}
		The formal definitions of FT gates \cite{WAhmed_CICM15} along with Definition 1 can be utilized to formally represent the FT of the ASN gateway in terms of its failure events. We systematically present the formalization of the ASN gateway FT by starting from the fourth level, i.e., the formalization of B1 sub-FT:
		\begin{flushleft}
			\small {\texttt{\bf{Definition 2: }}}\small{\texttt{$\forall$p t D1 D4 E1 E2 E3 E4 E5 E6 E7 E8 E9 E10 E21.\\
				B1\_FT p t D1 D4 E1 E2 E3 E4 E5 E6 E7 E8 E9 E10 E21 =\\
				 (OR [OR [atomic (fail\_event p D1 t);\\ \qquad \ \  \qquad
				AND [OR (gate\_list (fail\_event\_list p [E1; E2] t));\\ \qquad \ \  \qquad
				atomic (fail\_event p E21 t)];\\ \qquad \ \  \qquad
				OR (gate\_list (fail\_event\_list p [E3; E4; E5] t))];\\ \qquad
				OR [atomic (fail\_event p D4 t);\\ \qquad  \qquad
				AND [OR (gate\_list (fail\_event\_list p [E6; E7] t));\\ \qquad  \qquad
				atomic (fail\_event p E21 t)];\\ \qquad  \qquad
				OR (gate\_list (fail\_event\_list p [E8; E9; E10] t))]])
			}}
		\end{flushleft}
		
\noindent Where the random variables $D1$, $D4$, $E1-E10$ and $E21$ model the time-to-failure of the communication process ASN to RTDS. The diagram of B1 FT is similar to B2 FT, which can be seen in Table \ref{table:ASN_FT}. Additionally, the cut-set failure events in the above definition is already minimal, i.e., there are no combination of redundant failure events to be removed \cite{kornecki2013fault}. Therefore, the cut-sets and MCS for B1 sub-FT, in this case, are equivalent.

Similarly, other sub-FTs, such as B2-FT, A-FT, RT-FT and Internal-FT, which are at the fourth, third and second level of the ASN gateway FT can be formalized in HOL4 as shown in Table \ref{table:ASN_FT}. It is important to note that the formal definition of the top level or first level FT, in Table 3, builds upon the formal definitions of all the other sub-FTs and models the complete ASN gateway FT.

\begin{table}
	\caption{ASN Gateway FT Levels with their HOL Formalizations}
	\centering
	\scalebox{0.7}{
		\begin{tabular}{|l|l|}
			\hline
			ASN Sub-FTs &Formal Definitions of Sub-FTs in HOL \\
			\hline
			\hline
			\includegraphics[valign=t,scale=0.14]{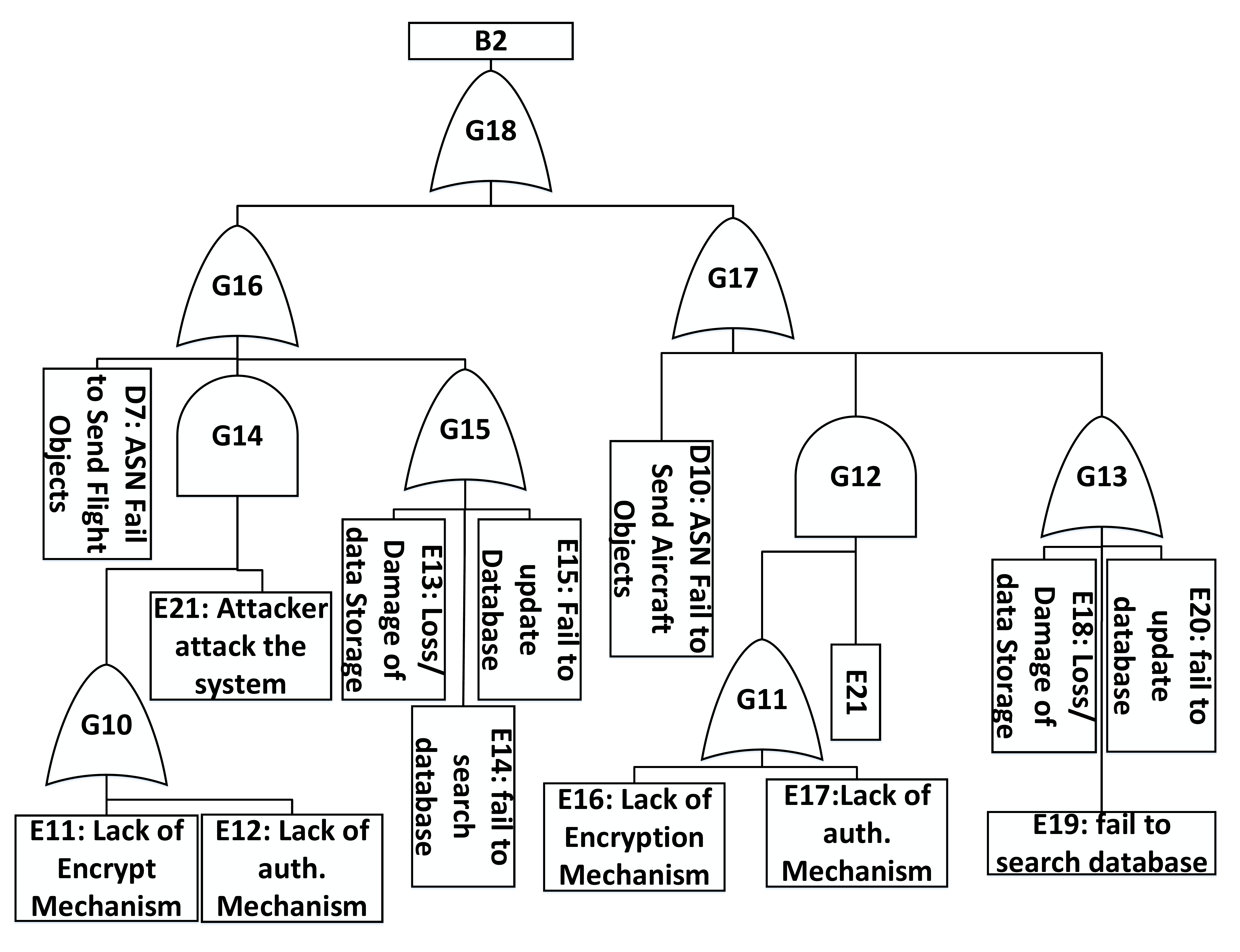}
			&
		$\!\begin{aligned}[t]
			& \texttt{($\mathbf{B2\_FT}$ p t D7 D10 E11 E12 E13 E14 E15}\\& \texttt{E16 E17 E18 E19 E20 E21)} =  \\
			& \texttt{OR [OR [atomic (fail\_event p D7 t);}\\
			&   \ \  \texttt{AND [OR (gate\_list}\\
            & \ \ \ \ \texttt{ (fail\_event\_list p [E11; E12] t));}\\
			& \ \ \ \texttt{atomic (fail\_event p E21 t)];}\\
			& \ \ \texttt{OR (gate\_list}\\
            & \ \ \ \ \texttt{ (fail\_event\_list p [E13; E14; E15] t))];}\\
			& \ \ \texttt{OR [atomic (fail\_event p D10 t);}\\
			& \ \ \texttt{AND [OR (gate\_list}\\
            & \ \ \ \ \texttt{ (fail\_event\_list p [E16; E17] t));}\\
			& \ \ \ \ \ \ \texttt{atomic (fail\_event p E21 t)];}\\
			& \ \  \texttt{OR (gate\_list}\\
            & \ \ \ \ \texttt{ (fail\_event\_list p [E18; E19; E20] t))]]}
        \end{aligned}$\\
			\hline
			\includegraphics[valign=t,scale=0.2]{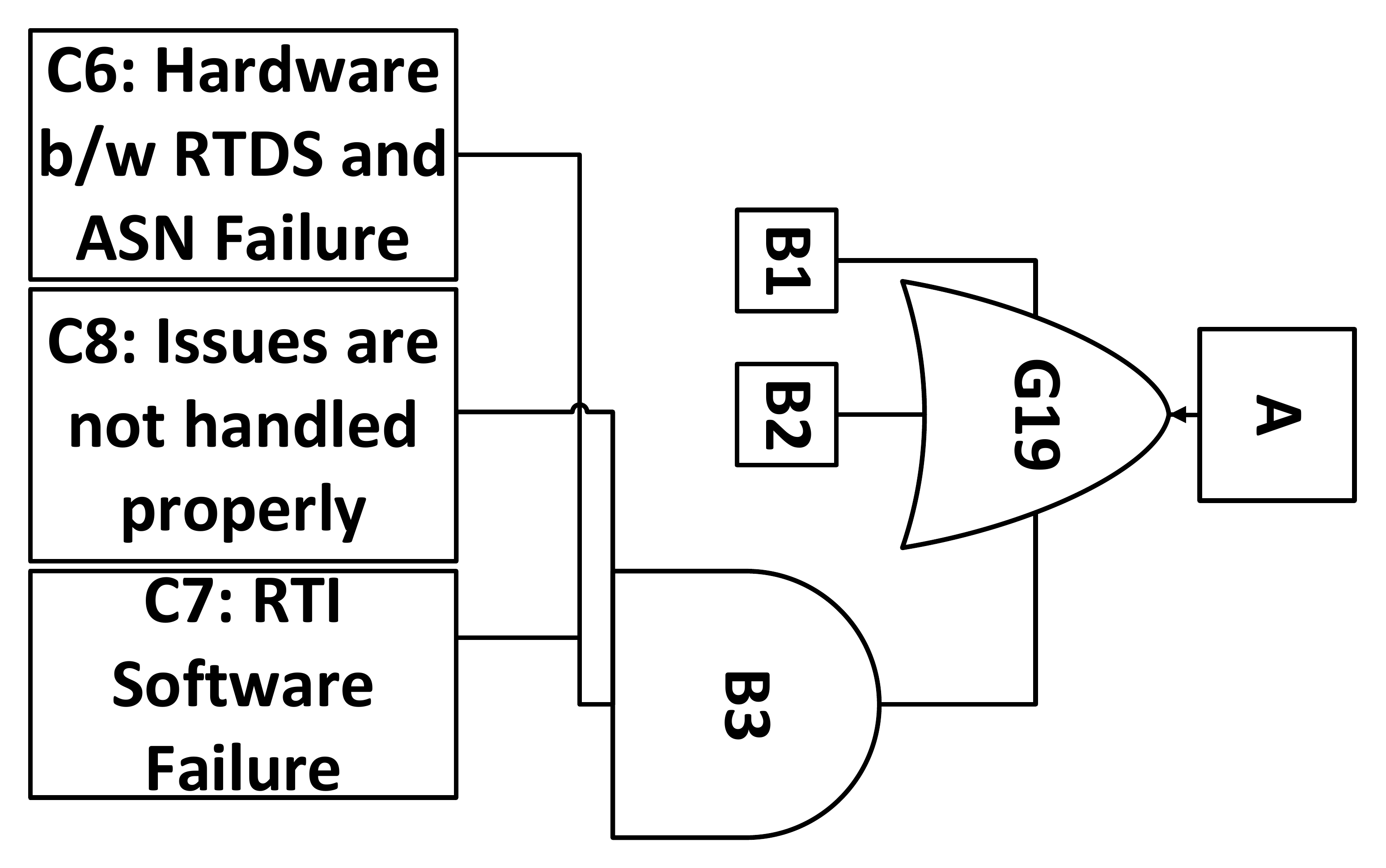}   &
        $\!\begin{aligned}[t]
			& \texttt{$\mathbf{A\_FT}$ p t D1 D4 D7 D10 E1 E2 E3 E4 E5 E6}\\
			& \ \ \  \texttt{E7 E8 E9 E10 E11 E12 E13 E14 E15 E16}\\
			&\ \texttt{  E17 E18 E19 E20 E21 C5 C6 C7 C8} =  \\
			& \texttt{OR [B1\_FT p t D1 D4 E1 E2 E3 E4 E5 E6 E7}\\
            & \texttt{\qquad E8 E9 E10 E21;}\\
			& \texttt{ B2\_FT p t D7 D10 E11 E12 E13 E14 E15 E16 E17}\\
            & \texttt{\qquad E18 E19 E20 E21;}\\
			& \  \texttt{ AND [OR (gate\_list}\\
            & \ \ \ \ \texttt{(fail\_event\_list p [C5; C6; C7] t));}\\
            & \ \ \ \texttt{atomic (fail\_event p C8 t)]]}
        \end{aligned}$\\
			\hline
			\includegraphics[valign=t,scale=0.2,angle= -90]{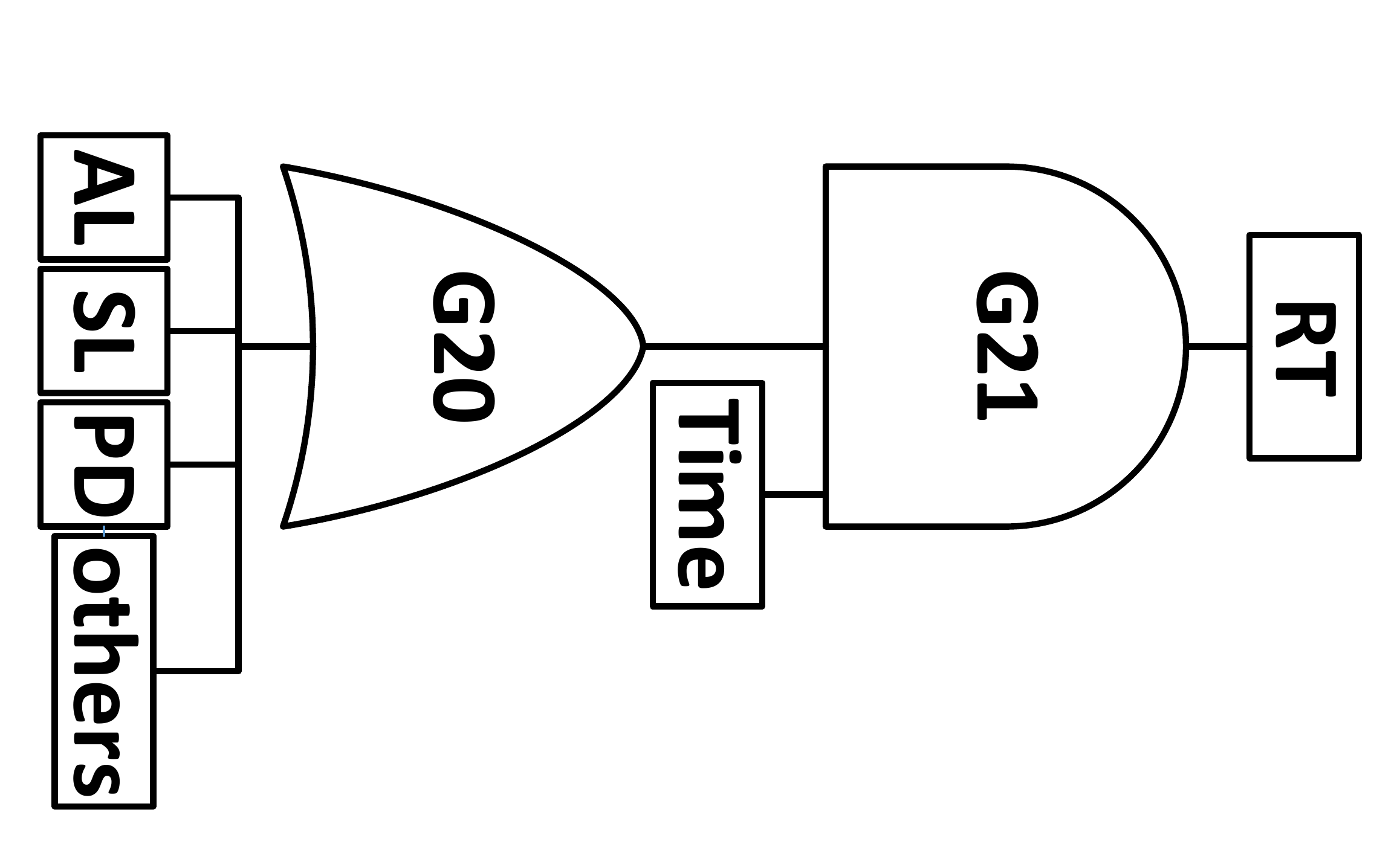}      &
        $\!\begin{aligned}[t]
			& \texttt{$\mathbf{RT\_FT}$ p t AL SL PD Others time} =  \\
			& \texttt{OR\_FT\_gate [B1\_FT p t D1 D4 E1 E2 E3 E4 E5 E6}\\
            & \texttt{E7 E8 E9 E10 E21;}\\
			& \texttt{ AND [OR (gate\_list (fail\_event\_list p}\\
            & \ \ \  \texttt{[AL; SL; PD; Others] t));}\\
            &  \  \texttt{ atomic (fail\_event p time t)]}
        \end{aligned}$ \\
			\hline
			\includegraphics[valign=t,scale=0.2,angle = -90]{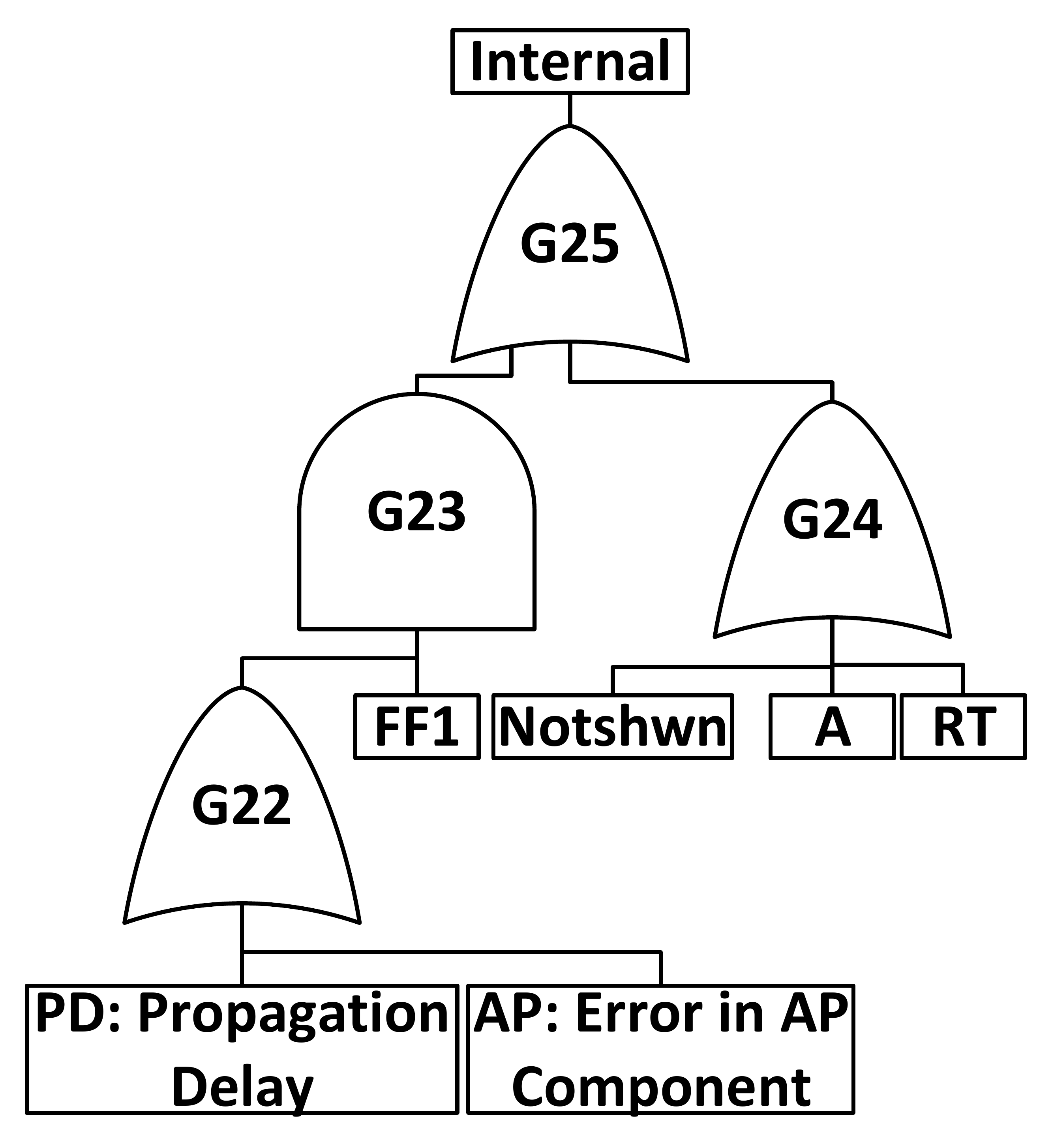}  &
        $\!\begin{aligned}[t]
			& \texttt{$\mathbf{Internal\_FT}$ p t FD AP FF1 D1 D4}\\
            & \texttt{D7 D10 E1 E2 E3 E4 E5 E6 E7 E8 E9 E10 E11 E12}\\
			& \  \ \texttt{E13 E14 E15 E16 E17 E18 E19 E20 E21 C5 C6}\\
            & \texttt{C7 C8	notshw AL SL PD Others time} =  \\
			& \texttt{OR [AND [OR (gate\_list}\\
            & \texttt{\qquad (fail\_event\_list p [FD; AP] t));}\\
            & \texttt{\qquad \ atomic (fail\_event p FF1 t)];}\\
			& \texttt{OR [A\_FT p t D1 D4 D7 D10 E1 E2 E3}\\
            & \texttt{\qquad E4 E5 E6 E7 E8 E9 E10}\\
			& \texttt{\ \qquad E11 E12 E13 E14 E15 E16 E17 E18}\\
            & \texttt{\qquad \ \ E19 E20 E21 C5 C6 C7 C8; notshw;}\\
			& \texttt{RT\_FT p t AL SL PD Others time]]}
		\end{aligned}$\\
			\hline
			\includegraphics[valign=t,scale=0.2]{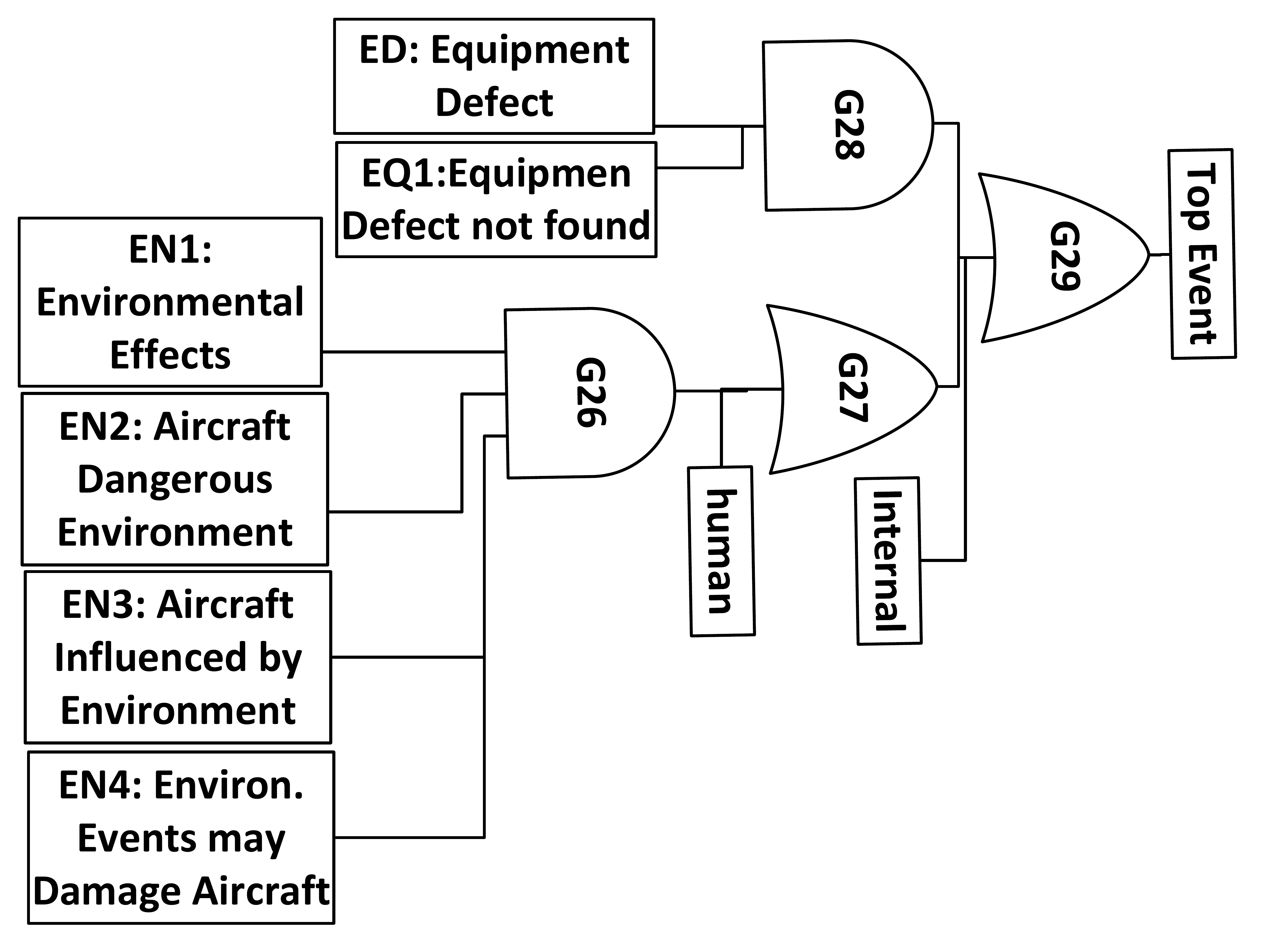}  &
        $\!\begin{aligned}[t]
			& \texttt{$\mathbf{ASN\_gateway\_FT}$ p t FD AP FF1 D1 D4}\\
			& \  \texttt{ D7 D10 E1 E2 E3 E4 E5 E6 E7 E8 E9 E10 E11 E12   }\\
			& \  \ \texttt{E13 E14 E15 E16 E17 E18 E19 E20 E21 C5 C6 C7}\\
            & \texttt{ C8 notshw AL SL PD Others time ED}\\
            & \texttt{ EQ1 EN1 EN2 EN3 EN4 human} =  \\
			& \texttt{OR [AND [OR (gate\_list}\\
            & \ \ \ \ \texttt{ (fail\_event\_list p [FD; AP] t));}\\
            & \texttt{\quad \quad  atomic (fail\_event p FF1 t)];}\\
			& \texttt{AND [OR [AND (gate\_list}\\
            & \texttt{\qquad (fail\_event\_list p [ED; EQ1] t));}\\
			& \texttt{OR [AND(gate\_list}\\
            & \texttt{\qquad (fail\_event\_list p [EN1; EN2; EN3; EN4] t));}\\
			& \ \texttt{\qquad \ fail\_event p human t];}\\
			& \texttt{Internal\_FT\_gate p t FD AP FF1 D1 D4 D7  }\\
            & \texttt{D10 E1 E2 E3 E4 E5 E6 E7 E8 E9 E10 E11 E12  }\\
			& \texttt{E13 E14 E15 E16  E17 E18 E19 E20 E21 }\\
			&\ \ \  \texttt{C5 C6 C7 C8 notshw AL SL PD Others time]]}\\
		\end{aligned}$\\
			\hline
		\end{tabular}} \label{table:ASN_FT}
	\end{table}
	


We consider that the random variables, associated with the failure events of the ASN gateway FT, exhibit the exponential distribution:
\begin{flushleft}
	\small{\texttt{\bf{Definition 3: }} \label{Exponential_distribution_def}
		\vspace{1pt} \small\texttt{$\vdash$ $\forall$ p X l. exp\_dist p X l = \\ \ \ \  $\forall$ x.  (CDF p X x =  if 0 $\leq$ x then 1 - exp (-l * x) else 0)
		}}
	\end{flushleft}
	
	\noindent The function \texttt{exp\_dist} guarantees that the CDF of the random variable $X$ is that of an exponential random variable with a failure rate $l$ in a probability space $p$. We classify a list of exponentially distributed random variables as follows:
	\begin{flushleft}
		\small{\texttt{\bf{Definition 4: }} \label{list_of exponential_distribution_function_def}
			\vspace{1pt} \small{\texttt{$\vdash$ $\forall$p L. list\_exp p [] L = T $\wedge$\\
					$\forall$p h t L. list\_exp p (h::t) L = exp\_dist p (HD L) h $\wedge$ list\_exp p t (TL L)
				}}}
			\end{flushleft}
			
			\noindent The function \texttt{list\_exp} accepts a list of failure rates, a list of random variables $L$ and
			a probability space $p$. It guarantees that all elements of the list $L$ are exponentially distributed with the corresponding failure rates, given in the other list, within the probability space $p$. For this purpose, it utilizes the list functions \texttt{HD} and \texttt{TL}, which return the \emph{head} and \emph{tail} of a list, respectively.

\subsection{Failure Assessment of NextGen ASN Gateway System}
  We now present the formal verification of all the sub-FTs, such as B1-FT, B2-FT, A-FT, RT-FT and Internal-FT. The formally verified results of these sub-FTs are then used to reason about the failure probability of overall ASN gateway communication system. Using the closed form expression of parallel-series RBD configuration, given in Equation (\ref{eq7:parallel-seriesRBD}), the failure probability of the B1-FT can be expressed mathematically  as follows:

\small{
\begin{equation}\label{B1_FT conclusion}
\begin {split}
&F_{B1}(t)=(1-e^{-(c_1+c_2+c_3+c_4)t})*(1-(1-e^{-C_{E1} t})(1-e^{-C_{E21} t})) (1 - (1-e^{-C_{E2} t})\\ & (1-e^{-C_{E21} t}))
 (1 - (1-e^{-C_{E6} t})(1-e^{-C_{E21} t})) (1 - (1-e^{-C_{E7} t})(1-e^{-C_{E21} t}))\\
\end{split}
\end{equation}}
\normalsize

To verify Equation (\ref{B1_FT conclusion}), we first verify a lemma that transforms the B1 sub-FT to its equivalent parallel-series RBD model as follow:

\begin{flushleft}
	\small {\texttt{\bf{Lemma 1: }}} \label{B1_eq_RBD}
	\vspace{1pt} \small{\texttt{$\vdash$ $\forall$ p t D1 D4 E1 E2 E3 E4 E5 E6 E7 E8 E9 E10 E21.\newline
	FTree p (B1\_FT p t D1 D4 E1 E2 E3 E4 E5 E6 E7 E8 E9 E10 E21) =\\ (rbd\_struct p ((parallel of\\ \
	($\lambda$a. series (rbd\_list (fail\_event\_list a)))) [[D1];[D4];[E1;E21];\\ \ \ \ [E2;E21]; [E3];[E4];[E5];[E6;E21];[E7;E21];[E8];[E9];[E10]]))
		}}
	\end{flushleft}

 Now, using the formal definition of B1-FT and Lemma 1, the failure probability of B1 sub-FT can be verified in HOL4 as follows:

\begin{flushleft}
	\small {\texttt{\bf{Theorem 5: }}} \label{series_connected_system_THM}
	\vspace{1pt} \small{\texttt{$\vdash$ $\forall$ p t D1 D4 E1 E2 E3 E4 E5 E6 E7 E8 E9 E10 E21 C\_E1 C\_E2\\ C\_E6
			C\_E7 C\_D1 C\_D4 C\_E3 C\_E4 C\_E5 C\_E8 C\_E9 C\_E10 C\_21.\newline
			    time\_positive t $\wedge$
			       prob\_space p $\wedge$ \newline
			in\_events p (fail\_event\_list p [D1;D4;E1;$\cdots$;E10;E21] t) $\wedge$ \newline
			 mutual\_indep p (fail\_event\_list p [D1;D4;E1;$\cdots$;E10;E21] t) $\wedge$ \newline
			 list\_exp p [C\_D1;C\_D4;C\_E1;$\cdots$;C\_E10;C\_E21]  [D1;D4;E1;$\cdots$;E10;E21] $\Rightarrow$ \\
			\ (prob p (\textbf{B1\_FT }p t D1 D4 E1 E2 E3 E4 E5 E6 E7 E8 E9 E10 E21) =\\
			\  \     1 - exp(-(t * list\_sum [C\_D1;C\_D4;C\_E3;C\_E4;C\_E5;C\_E8;C\_E9;C\_E10])) *\\
			 \ \ \  list\_prod(one\_minus\_exp\_prod t\\ \ \ \ \
			 [[C\_E1;C\_E21];[C\_E2;C\_E21];[C\_E6;C\_E21];[C\_E7;C\_E21]]))
		}}
	\end{flushleft}
	\noindent The  function \texttt{exp} represents the exponential function. The function \texttt{list\_sum} is used to sum all the elements of the given list of failure rates, the function \texttt{one\_minus\_exp} accepts a list of failure rates and returns a one minus list of exponentials and the function \texttt{one\_minus\_exp\_prod}  accepts a two dimensional list of  failure rates and returns a list with one minus product of one minus exponentials of every sub-list. For example, \texttt{one\_minus\_exp\_prod}$[[c1; c2; c3]; [c4; c5]; [c6; c7; c8]]$ $ x = [1 - ((1 - e^ {-(c1)x})*(1 - e^{ -(c2)x})*(1 - e^{ -(c3)x}));$
	$(1 - (1 - e^{ -(c4)x}) * (1 - e^ {-(c5)x})); (1 - (1 - e^ {-(c6)x}) * (1 - e^{ -(c7)x}) *(1 - e^ {-(c8)x})) ]$. The first assumption ensures that the variable \texttt{t} models time $t$ as it can acquire positive integer values only. The next assumption ensures that \texttt{p} is a valid probability space based on the probability theory in HOL \cite{mhamdi_11}. The next two assumptions ensure that the events corresponding to the failures modeled by the random variables \texttt{D1, D2, E1} to \texttt{E10} and \texttt{E21} are valid events from the probability space \texttt{p} and they are mutually independent. Finally, the last assumption characterizes the random variables \texttt{D1, D2, E1} to \texttt{E10} and \texttt{E21}, as exponential random variables with failure rates \texttt{C\_D1, C\_D2, C\_E1} to \texttt{C\_E10} and \texttt{C\_E21}, respectively. The conclusion of Theorem 5 represents the failure probability of the communication process between ASN to RTDS in terms of the failure rates of the components involved during the communication process.
	The proof of Theorem 5 is primarily based on Theorem 4 and some fundamental facts and axioms of probability.
	
	Similarly, the failure probabilities of other sub-FTs, i.e., B1-FT, B2-FT, A-FT, RT-FT and Internal-FT, are verified in HOL4 \cite{waqar_ASN_gateway_15}. These theorems are verified under the same assumptions as the one used in Theorem 5.
	
	Now, using the formal definitions of ASN gateway sub-FTs, given in Table \ref{table:ASN_FT}, and their verified failure probability results \cite{waqar_ASN_gateway_15}, we formally verified the failure probability of the complete ASN gateway system as follows:

	\begin{flushleft}
		\small {\texttt{\bf{Theorem 6: }}} \label{ASN_gatway_THM}
		\vspace{1pt} \small{\texttt{$\vdash$
				(prob p (\textbf{ASN\_gateway\_FT} p t FD AP FF1 D1 D4 D7 D10 E1 $\cdots$ E21 C5 C6 C7 C8 notshw AL SL PD Others time ED EQ1 EN1 $\cdots$ EN4 human) =\\
				\ \   1 - (list\_prod(one\_minus\_exp\_prod t [[C\_ED;C\_EQ1];\\
				\ \ \ \ \qquad [C\_EN1;C\_EN2;C\_EN3;C\_EN4];[C\_E6;C\_E21]])) *\\
				\ \	   exp (-(t*C\_human)) * exp -(t*C\_notshw) *\\
				\ \ 1 - (list\_prod(one\_minus\_exp\_prod t [[C\_FD;C\_FF1];[C\_AP;C\_FF1]]) * \\
				\ \     1 - (1 - exp(-(t*list\_sum
				[C\_D1;C\_D4;C\_E3;C\_E4;C\_E5;C\_E8;C\_E9;C\_E10])) *\\
				\ \     list\_prod(one\_minus\_exp\_prod t
				[[C\_E1;C\_E21];[C\_E2;C\_E21];\\
				\ \ \ \     [C\_E6;C\_E21];[C\_E7;C\_E21]])))*\\
				\ \   1 - exp(-(t*list\_sum[C\_D7;C\_D10;
				C\_E13;C\_E14;C\_E15;C\_E18;C\_E19;C\_E20])) *\\
				\ \    list\_prod(one\_minus\_exp\_prod t\\
				\ \ \ \ \  [[C\_E11;C\_E21];[C\_E12;C\_E21];[C\_E16;C\_E21];[C\_E17;C\_E21]])) *\\
				\ \    list\_prod(one\_minus\_exp\_prod t
				[[C\_C5;C\_C8];\\
				\ \ \ \ \ [C\_C6;C\_C8];[C\_C7;C\_C8]]))))))*\\
				\ \    list\_prod(one\_minus\_exp\_prod t
				[[C\_AL;C\_time];\\
				\ \ \ \ [C\_SL;C\_time];[C\_PD;C\_time];
				[C\_other;C\_time]]))))
			}}
		\end{flushleft}	
		
		\noindent The assumptions of the above theorem are similar to the ones used in Theorem 5 and its proof is based on Theorem 4 and some basic arithmetic lemmas and probability theory axioms. The proof of Theorems 5  and 6  and the formalization of sub-FTs, presented in Table \ref{table:ASN_FT},  with their corresponding probability of failure took more than 2500 lines of HOL codes \cite{waqar_ASN_gateway_15} and about 125 man-hours.
	
In order to facilitate the use of our formally verified results by industrial design engineers for their failure analysis, we have also developed a set of SML scripts to automate the simplification step of these theorems for any given failure rate list corresponding to the NextGen ATM system components. For instance, the output of the  \texttt{auto\_ASN\_gateway\_FT} script \cite{waqar_ASN_gateway_15} for the automatic simplification of Theorem 6 is as follows:

\begin{flushleft}
		\vspace{1pt} \small{\texttt{$\vdash$
				(prob p (\textbf{ASN\_gateway\_FT} p t FD AP FF1 D1 D4 D7 D10 E1 $\cdots$ E21 C5 C6 C7 C8 notshw AL SL PD Others time ED EQ1 EN1 $\cdots$ EN4 human) =\\
				$\mathit{1 -
(1 - (1 - e^{(-5 / 2)}) * (1 - e^{(-3 / 2)})) *
((1 -
  (1 - e^{(-1 / 2)}) *
  ((1 - e^{(-2)}) *}$\\$\mathit{ ((1 - e^{(-3 / 2)}) * (1 - e^{(-4)})))) *
 e^{(-9 / 2)}) *
((1 - (1 - e^{(-7 / 2)}) * (1 - e^{(-3)})) *}$\\ $\mathit{
 (1 - (1 - e^{(-4)}) * (1 - e^{(-3)})) *
 (e^{(-4)} *
  ((1 - (1 - e^{(-1 / 2)}) * (1 - e^{(-3)})) *}$\\ $\mathit{
   ((1 - (1 - e^{(-1 / 2)}) * (1 - e^{(-3)})) *
    ((1 - (1 - e^{(-1 / 2)}) * (1 - e^{(-3)})) *}$\\ $\mathit{
     (1 - (1 - e^{(-1 / 2)}) * (1 - e^{(-3)}))))) *
  (e^{(-321 / 20)} *
   ((1 - (1 - e^{(-1 / 2)}) * (1 - e^{(-3)})) *}$\\ $\mathit{
    ((1 - (1 - e^{(-1 / 2)}) * (1 - e^{(-3)})) *
     ((1 - (1 - e^{(-1 / 2)}) * (1 - e^{(-3)})) *}$\\ $\mathit{
      (1 - (1 - e^{(-1 / 2)}) * (1 - e^{(-3)})))))) *
  ((1 - (1 - e^{(-3 / 2)}) * (1 - e^{(-2)})) *}$\\ $\mathit{
   ((1 - (1 - e^{(-1 / 2)}) * (1 - e^{(-2)})) *
    (1 - (1 - e^{(-1 / 2)}) * (1 - e^{(-2)}))))) * e^{(-1)} *}$\\ $\mathit{
 ((1 - (1 - e^{(-7 / 2)}) * (1 - e^{(-3)})) *
  ((1 - (1 - e^{(-3 / 2)}) * (1 - e^{(-3)})) *}$\\ $\mathit{
   ((1 - (1 - e^{(-1 / 2)}) * (1 - e^{(-3)})) *
    (1 - (1 - e^{(-5 / 2)}) * (1 - e^{(-3)}))))))
			}$}}
		\end{flushleft}
	
\noindent With a very little modification, these kind of automation scripts can facilitate industrial design engineers to accurately determine the failure probabililty of many other safety-critical systems.
		
		\section{Conclusion}
		
		The accuracy of failure analysis is a dire need for safety and mission-critical applications, like the avionic ASN gateway communication system, where a slight error in the failure analysis may lead to disastrous situations including the death of innocent human lives or heavy financial setbacks. In this paper, we presented a deep embedding based formalization of commonly used FT gates, which facilitates the transformation of a FT model to its equivalent RBD model. The transformation considerably reduces the complexity of the FT analysis compared to our earlier FT formalization \cite{WAhmed_CICM15}.  For illustration, the paper presents the formalization of each level of ASN gateway FT and then building upon this formalization the failure probability of  overall ASN gateways communication system is verified.									

\bibliographystyle{splncs}
\bibliography{biblio}

\end{document}